\documentclass[a4paper,11pt]{article}
\usepackage{pos}
\usepackage{multicol}
\usepackage{float}
\usepackage{enumitem}
\usepackage{siunitx}
\usepackage{gensymb}
\usepackage{lineno}
\usepackage{wrapfig}
\usepackage[utf8]{inputenc}
\usepackage[T1]{fontenc}
\usepackage{subcaption}
\usepackage{tabularray}
\usepackage[bottom]{footmisc}
\usepackage{gensymb,multirow,wrapfig,lineno,enumitem}
\makeatletter
\newcommand*{\ssymbol}[1]{\ensuremath{\@fnsymbol{#1}}}
\makeatother

\title{Constraining the origin of the highest-energy cosmic-ray events detected by the Pierre Auger Observatory: a three-dimensional approach}
\ShortTitle{Constraining the origin of the highest-energy cosmic-ray events detected by Auger}

\author*[ab]{Marta Bianciotto}

\affiliation[a]{INFN-Sezione di Torino, Via Pietro Giuria 1, 10125, Torino, Italy}
\affiliation[b]{Physics department, Università degli Studi di Torino, Via Pietro Giuria 1, 10125, Torino, Italy}

\onbehalf{for the Pierre Auger Collaboration$^c$}
\affiliation[c]{Observatorio Pierre Auger, Av.\ San Mart{\'\i}n Norte 304, 5613 Malarg\"ue, Argentina\\
Full author list: {\rm\url{https://www.auger.org/archive/authors_icrc_2025.html}}}
\emailAdd{spokespersons@auger.org}

\abstract{Unveiling the sources of ultra-high-energy cosmic rays remains one of the main challenges of high-energy astrophysics. Measurements of anisotropies in their arrival directions are key to identifying their sources, yet magnetic deflections obscure direct associations. In this work, we reconstruct the sky regions of possible origin of the highest-energy cosmic-ray events detected by the Pierre Auger Observatory by tracing their trajectories through Galactic magnetic fields using up-to-date models, while fully accounting for energy and directional uncertainties. A mixed composition at injection is assumed to model the detected charge distributions of such events. Different classes of astrophysical sources are investigated and tested for a correlation with the inferred regions of origin of the events. By incorporating constraints on the maximum propagation distances, we also allow for a three-dimensional localization of the possible source regions. Our findings provide new constraints on the sources of the highest-energy cosmic particles and offer fresh insights into the role of Galactic magnetic fields in shaping the observed ultra-high-energy cosmic-ray sky.}

\ConferenceLogo{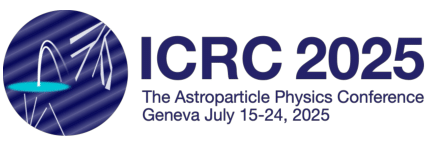}

\FullConference{39th International Cosmic Ray Conference (ICRC2025)\\
 15–24 July 2025\\
Geneva, Switzerland\\}

\begin{document}
\maketitle

\section{Introduction}
Ultra-high-energy cosmic rays (UHECRs) are the most energetic particles observed in nature, being defined as charged particles with energies exceeding $10^{18}\,\mathrm{eV}$. Their origin remains uncertain, as they are deflected by extragalactic and Galactic magnetic fields (EGMFs and GMFs) during propagation.
While intermediate-scale anisotropy measurements have provided hints of structure in their arrival directions, individual source association remains challenging.
Significant progress in the search for UHECR sources has been made thanks to the data from the Pierre Auger Observatory, the world’s largest area and exposure detector for UHECRs \cite{PierreAuger:2015eyc}.

% In this contribution, we present our work to reconstruct the possible sky regions of origin of UHECRs with energies $E \ge 100\,\mathrm{EeV}$ through Galactic backtracking, using \MB{up-to-date GMF models} \cite{Unger:2023lob}.
In this contribution, we present our work to constrain the potential origin of UHECRs with energies $E \ge 100\,\mathrm{EeV}$. We reconstruct their possible sky regions of origin via Galactic backtracking, using up-to-date GMF models \cite{Unger:2023lob}, and apply horizon constraints that limit the accessible source volume. This three-dimensional approach was first introduced in \cite{Bourriche:2023omi}, followed by \cite{Unger:2023hnu, uf_uhecr2024, Bourriche:2024bbe}.
Several classes of astrophysical sources are tested for a spatial correlation with the reconstructed sky regions. 

\section{Data set and source catalogs}
\label{sec:dataset}

Our study is based on the 40 most energetic events recorded at the Pierre Auger Observatory during Phase~I from Jan.~2004 to Dec.~2022. Of these, 36 were detected from 2004 to 2020 and were included in the catalog released in 2023 \cite{PierreAuger:2022qcg}, whereas the other 4 were detected in 2021 and 2022. As reported in \cite{PierreAuger:2022qcg}, the statistical uncertainty on the energy is of the order of $\sim$8\%, while the systematic uncertainty is assumed to be $\sim$14\%. At the highest energies, arrival directions are reconstructed with a precision better than $\sim0.4 \degree$.

Different astrophysical source classes are investigated and tested for correlations with the reconstructed UHECR source regions. In particular, we consider six catalogs of sources:
\begin{enumerate}[itemsep=0pt, parsep=0pt, topsep=0pt]
    \item 44 nearby galaxies with a high star formation rate, denoted as starburst galaxies (SBGs), based on the Lunardini catalog \cite{Lunardini:2019zcf} and weighted by their radio fluxes -- ``starburst galaxies'';
    \item 523 active galactic nuclei (AGNs), based on the Swift-BAT 105-month catalog \cite{Oh:2018wzc} and weighted by their hard X-ray fluxes -- ``all AGNs'';
    \item 26 jetted active galactic nuclei, based on the Fermi-LAT 3FHL catalog \cite{Fermi-LAT:2017sxy} and weighted by their $\gamma$-ray fluxes -- ``jetted AGNs'';
    \item $44{,}113$ galaxies of all types from the 2MASS catalog, based on the Two Micron All Sky Survey \cite{skrutskie2006two} and weighted by their near-infrared fluxes -- ``all galaxies'';
    \item 575 radio galaxies from the van Velzen catalog \cite{vanVelzen:2012fn} -- ``radio galaxies'';
    \item $\sim400{,}000$ galaxies of all types from the Biteau catalog \cite{Biteau:2021pru}, based on 2MASS, HyperLEDA, and Local Volume data, and weighted by their stellar mass (weight 1) or star formation rate (weight 2) estimates -- ``all galaxies B''.
\end{enumerate}
The Local Group galaxies (D $<1$~Mpc) are excluded from all catalogs. The first four catalogs are the same as those employed in \cite{PierreAuger:2022axr}.
%
%The Lunardini, Swift-BAT, Fermi-LAT and 2MASS catalogs used in this work are the same as those employed in previous Pierre Auger Collaboration intermediate-scale anisotropy studies, such as \cite{PierreAuger:2022axr}.
%The Van Velzen catalog was already employed in \cite{Unger:2023hnu}.

\section{Determination of the mass of the fragments at Earth}
\label{sec:masses}

Deflections of charged particles in magnetic fields depend on their rigidity, therefore it is necessary to model the probability distributions of the detected charge of the events. To this aim, we perform simulations with CRPropa 3.2 \cite{AlvesBatista:2022vem}. According to the results of the combined fit of the energy spectrum and mass composition measured at the Pierre Auger Observatory \cite{PierreAuger:2022atd}, we assume a mixed composition at the source, an injection spectrum $\propto E^{1.47}$ and a rigidity cutoff $\mathcal{R}_{\text{max}} = 10^{18.19}$~V. The events are generated uniformly in a distance range $\in [1,200]$~Mpc.

For each of the 40 events with~$E \ge 100\,\mathrm{EeV}$, we estimate their predicted mass composition at Earth by selecting simulated events whose energies match the observed ones, via a weight $w(E,E_i)$ \cite{Unger:2023hnu}. This is defined as:
\begin{equation}
    w(E,E_i) = e^{-\frac{(E - E_i)^2}{2\sigma^2}}
    \label{eq:weight}
\end{equation}
and it is proportional to the Gaussian probability to reconstruct the observed energy $E_i$ at Earth, given the simulated energy $E$ and the statistical uncertainty $\sigma \sim 8\%$.

\begin{figure}[H]
\centering
\includegraphics[width=0.55\linewidth]{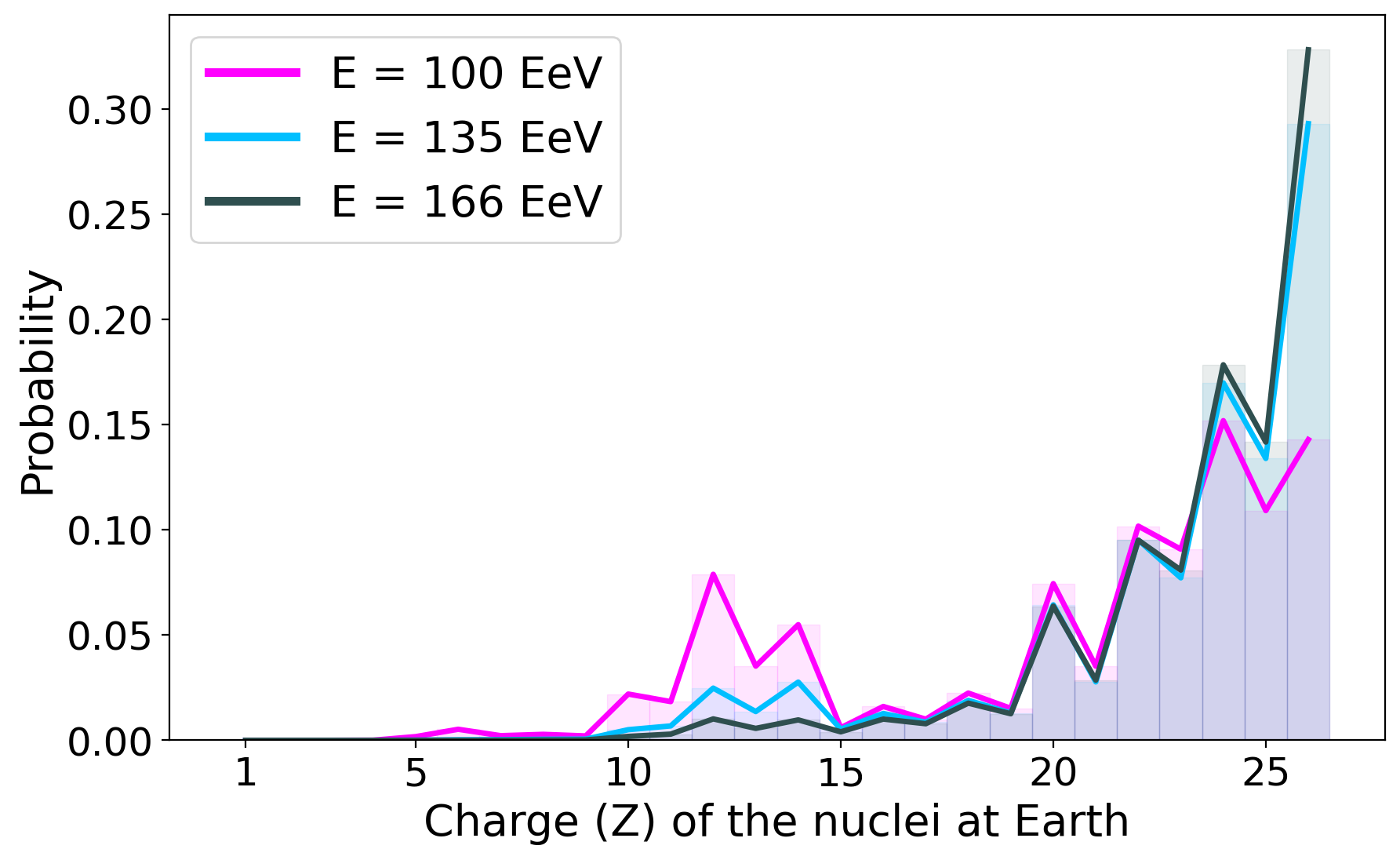}
\caption{Simulated charge distributions of fragments at Earth compatible with the detected energies of three events in the Auger dataset and assuming source composition as in~\cite{PierreAuger:2022atd}.}
\label{fig:mass}
\end{figure}

Fig.~\ref{fig:mass} illustrates the distributions of charges from $Z=1$ to $Z=26$ of the fragments at Earth with energies compatible with three events in the dataset. These events are detected with $E_1 = \SI{166}{EeV}$, $E_2 = \SI{135}{EeV}$ and $E_3 = \SI{100}{EeV}$.
At higher energies, the distribution is dominated by heavy elements, while going towards lower energies the CNO group contribution starts to be non-negligible.

\section{Determination of the UHECR horizon}
\label{sec:horizon}

The three-dimensionality of our analysis is only possible if we take into account the UHECR horizon -- the maximum distance within which the bulk of UHECRs with $E \geq 100$~EeV can be produced. The calculation is carried out using the same simulation setup described in Sec.~\ref{sec:masses}.

For each distance, we define the attenuation factor as the ratio of particles detected at Earth to particles injected in the same energy interval, similarly to \cite{Harari:2006uy, Globus:2022qcr, Unger:2023hnu}.
In particular, the attenuation factor for an event of energy $E_i$ and a source of distance $d$, is:
\begin{equation}
a(E_i,d) \propto
\sum_{A=1}^{56}\int dE\,w(E,E_i)\,T_A(E,d),
\label{eq:attenuation_window}
\end{equation}
where \(T_A(E,d)\) is the differential spectrum at Earth of particles with mass number \(A\) injected from sources at distance \(d\) and $w(E,E_i)$ is the weight defined in Eq.\,(\ref{eq:weight}). 
The attenuation factor is summed over all nuclear species that can be possibly detected at Earth.

\begin{figure}[!t]
\centering
\includegraphics[height=4.35cm]{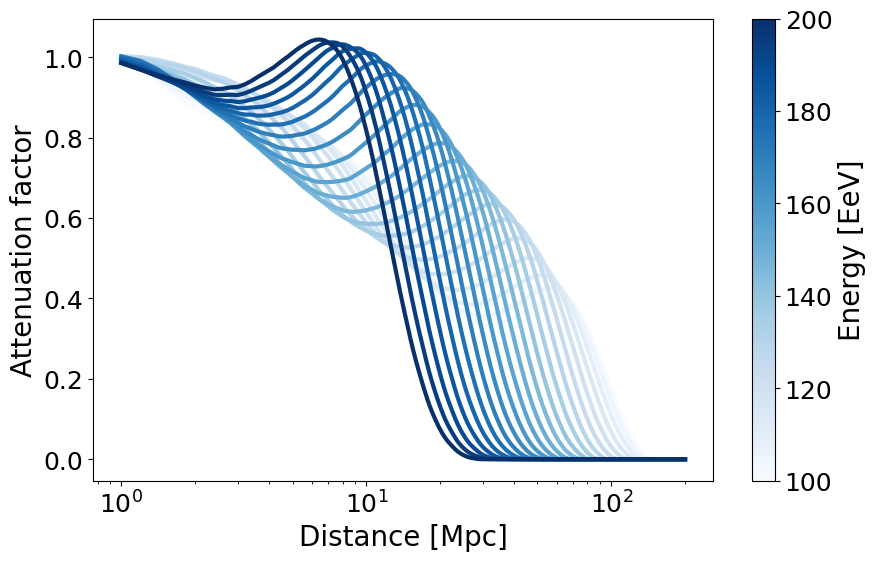}
\includegraphics[height=4.35cm]{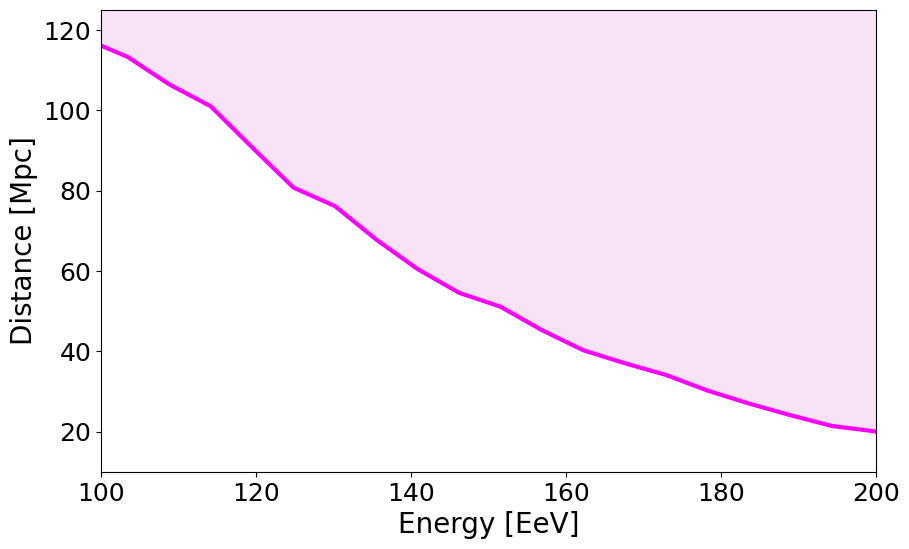}
\caption{\textbf{Left panel:} Attenuation factor \(a(E_i,d)\), as a function of the distance, for several energies (color scale). \textbf{Right panel:} Horizon distance \(D_{0.1}(E_i)\) obtained from the condition \(a(E_i,d) = 0.1\). Both panels adopt the combined-fit source model described in Sec.~\ref{sec:masses}.}
\label{fig:attenuation}
\end{figure}

In the left panel of Fig.~\ref{fig:attenuation}, the attenuation factor is shown as a function of the distance, for several representative energies.
In particular, $a(E_i, d) = 1$
means no attenuation, and $a(E_i,d) = 0$ means complete attenuation.
In the right panel, the corresponding distance \(D_{0.1}(E_i)\), defined by the condition \(a(E_i,d)=0.1\), is displayed. We assume this value as the maximum distance that a cosmic ray of energy $E_i$ is allowed to travel, hence the horizon.

% \newpage
\section{Galactic backtracking}
\label{sec:backtracking}

The sky regions of possible origin of the 40 UHECR events detected by the Pierre Auger Observatory are reconstructed through Galactic backtracking using CRPropa 3.2. This approach consists in simulating the propagation of a particle of same rigidity but opposite charge, starting from the Earth and propagating it backward through the Galactic magnetic field.
As a result, the backward-propagated antiparticle traces a trajectory equivalent to the forward path of the original cosmic ray.
%, thus providing an estimate of the cosmic ray incoming direction prior to Galactic magnetic field deflections.
%
The trajectory is followed until the particle exits the Galaxy, which is modeled as a sphere of 20~kpc radius. The final direction of the particle at this boundary is taken as the best estimate of the source direction when extragalactic magnetic fields are neglected.
In our analysis, %the most up-to-date Galactic magnetic field models are employed. 
the UF23 ensemble of eight coherent model variations \cite{Unger:2023lob} is used to model the regular field and the JF12 model with corrections by the Planck Collaboration \cite{Planck:2016gdp} to model the turbulent field. Extragalactic magnetic field deflections are neglected.

To account for statistical uncertainties in the detected energy and arrival direction of the particle, the randomness of the turbulent component of the Galactic magnetic field, and the possible particle masses, the backtracking procedure is repeated $20{,}000$ times. In each iteration, the energy of the particle is drawn from a Gaussian distribution centered on the nominal value, with a standard deviation of $8\%$. Its Galactic longitude and latitude are also sampled from a Gaussian distribution centered on the nominal arrival direction, with a standard deviation of $0.5\degree$. The charge is determined from the probability density functions described in Sec.~\ref{sec:masses} (see, e.g., Fig.~\ref{fig:mass}). The turbulent magnetic field is generated for each iteration with a different seed extracted from a uniform distribution.
%The seed for the random component of the Galactic magnetic field is extracted from a uniform distribution. %A similar approach was employed in \cite{Unger:2023hnu}.

\begin{figure}[H]
    \centering
    \includegraphics[width=0.48\linewidth]{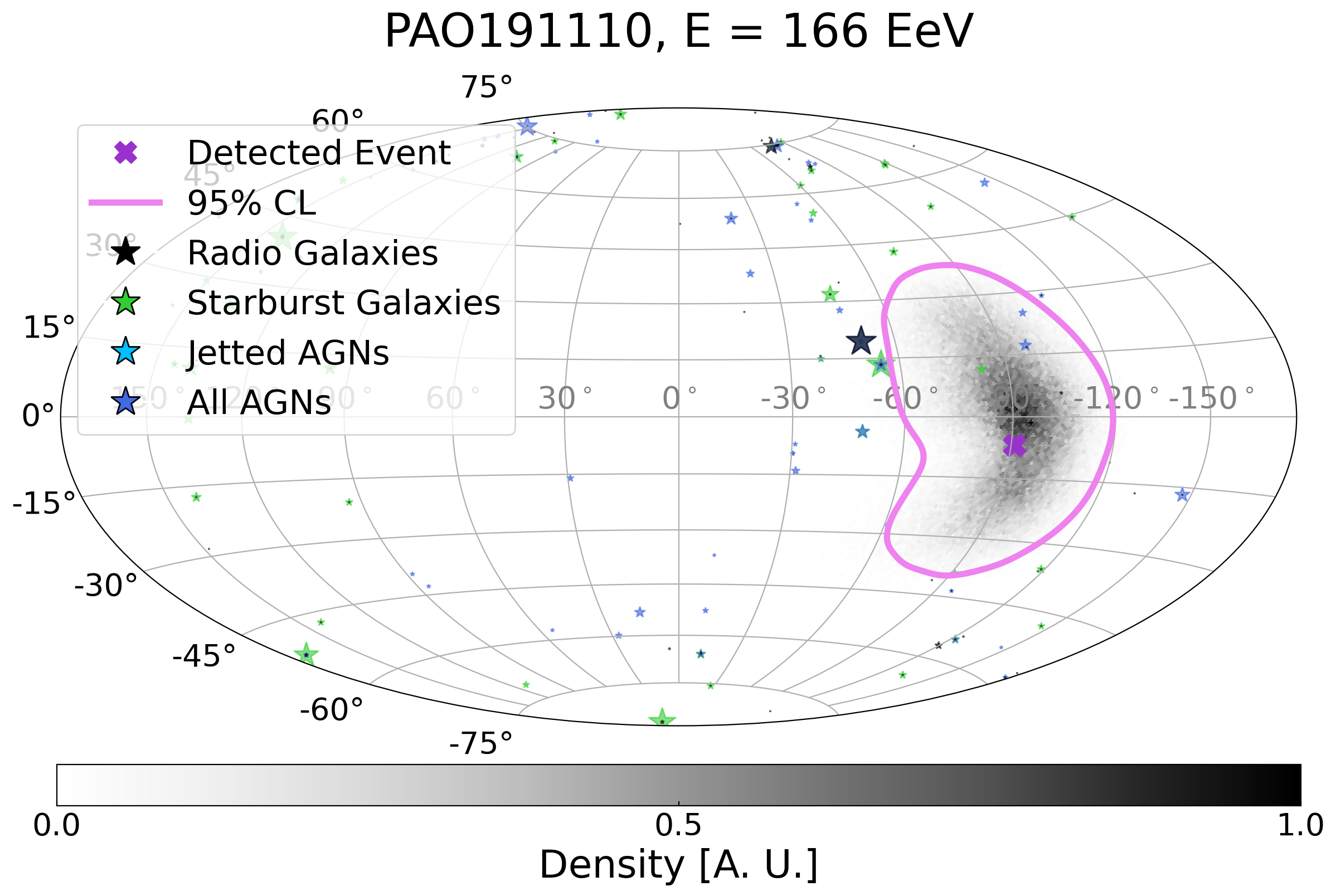}
    \includegraphics[width=0.48\linewidth]{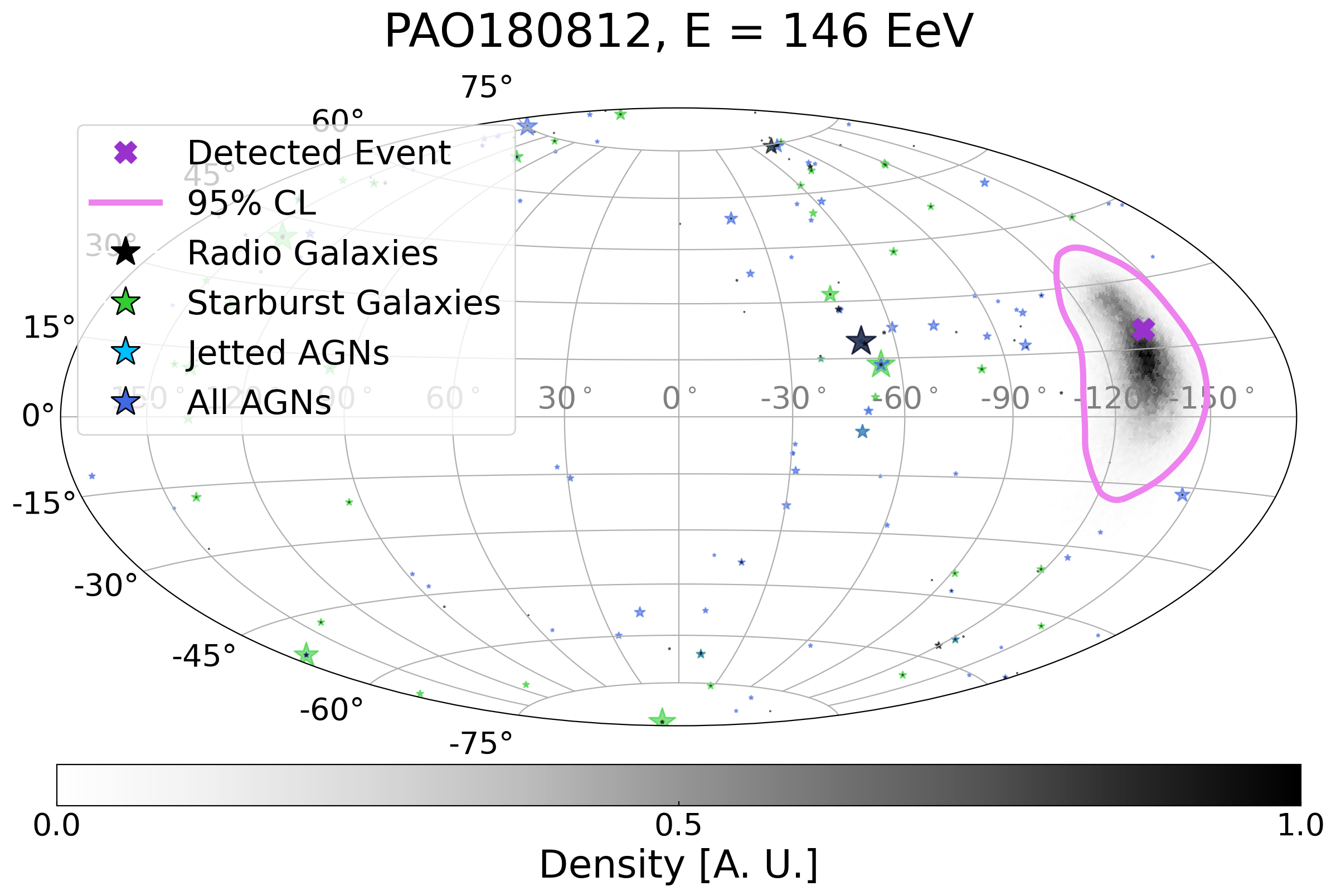}
    \caption{\small The combination of $20{,}000$ backtracking iterations for the eight model variations is shown in a \textit{healpy} map with $\mathit{nside}=64$, in Galactic coordinates, for candidate host galaxies for the two events \texttt{PAO191110} (\textbf{left panel}) and \texttt{PAO180812} (\textbf{right panel}).
    The violet line represents the 95\% confidence level (CL) that identifies the inferred region of origin of the event. 
    Only the sources that satisfy the maximum distance criterion of Sec.~\ref{sec:horizon} are shown in the maps.
    }
    \label{fig:CL}
\end{figure}

In Fig.~\ref{fig:CL}, the sky regions of origin inferred employing this method are shown for two events with energies above 100~EeV. 
The maps result from the combination of the eight GMF model variations of the UF23 ensemble.
The gray scale represents the pixel density, namely $\rho = \sum_{j=1}^8\rho_{j}$, where the index $j$ runs over the models and $\rho$ is normalized to the maximum.
All the 40 events are studied using this approach. Their localization uncertainties vary from $\sim$3\% to $\sim$33\% of $4\pi$, depending on the latitudes and energies of the events. 

\section{Correlation with astrophysical sources}
\label{sec:results}

In this Section, we investigate possible correlations between the inferred regions of localization of the 40 events above 100~EeV detected at the Pierre Auger Observatory and several classes of astrophysical sources.

\begin{figure}[!t]
\centering
\includegraphics[width=0.48\linewidth]{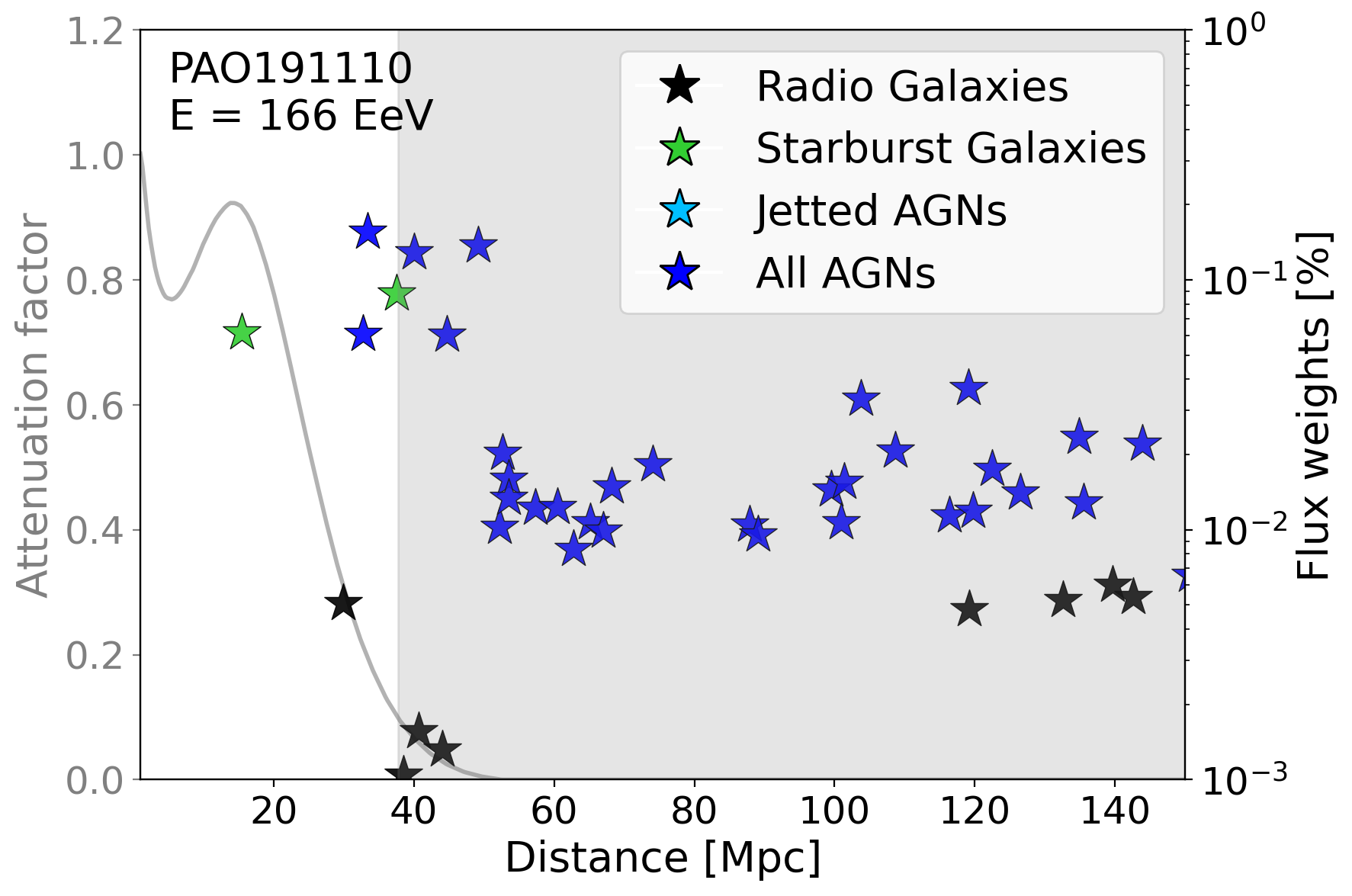}
\includegraphics[width=0.48\linewidth]{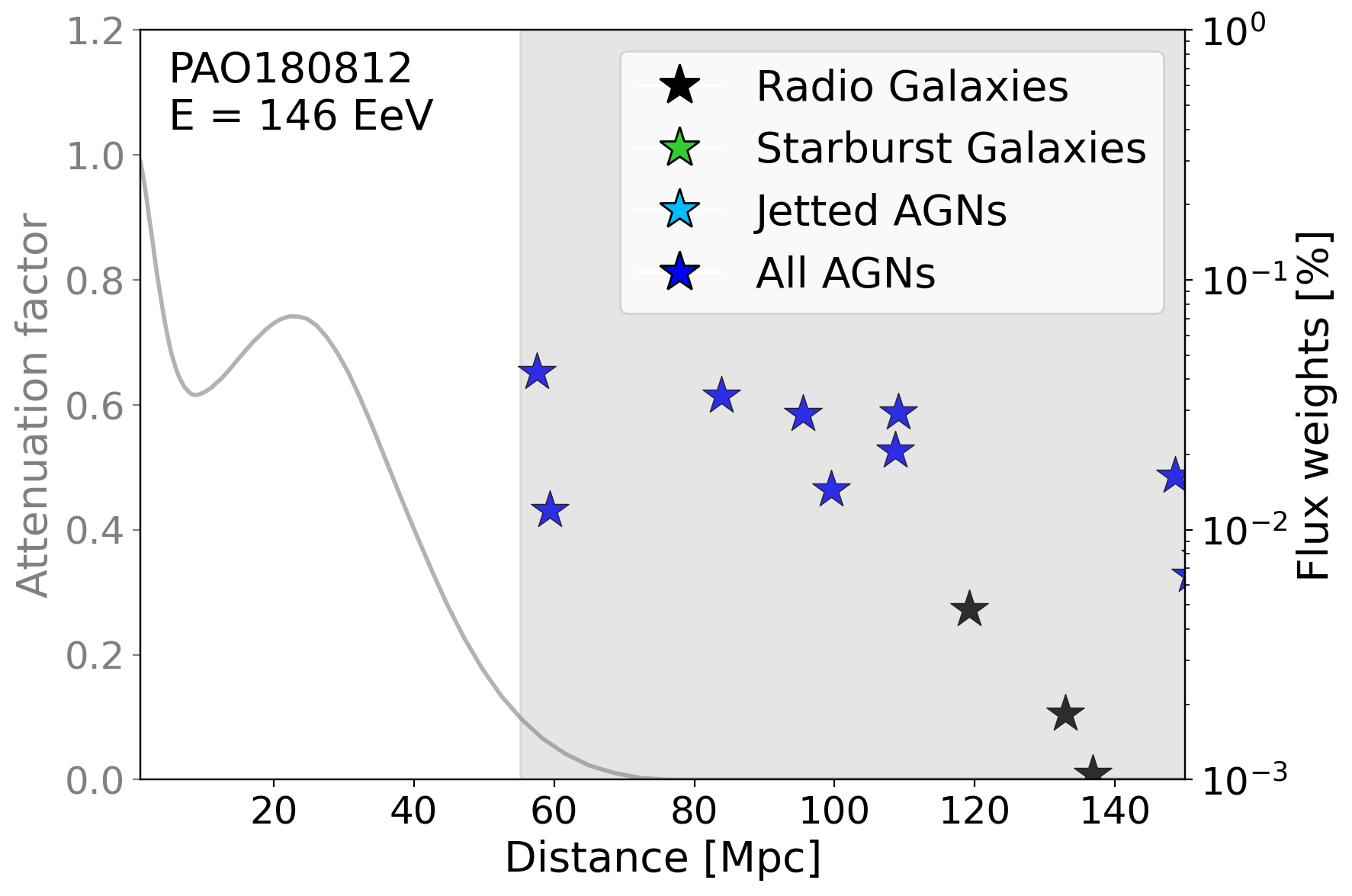}
\caption{\small Candidate host galaxies for the two events \texttt{PAO191110} (\textbf{left panel}) and \texttt{PAO180812} (\textbf{right panel}).  
Only galaxies whose positions fall inside the 95\% CL region (Sec.~\ref{sec:backtracking}) are shown; marker colors distinguish the source catalogs.  
\textit{Left ordinate} (solid grey curve): attenuation factor \(a(E,d)\) from Sec.~\ref{sec:horizon}. The shaded gray band marks distances where \(a<0.1\) and are therefore excluded by our horizon criterion.  
\textit{Right ordinate} (colored symbols): relative flux weight of each galaxy, normalized to the maximum flux in the sample.}
\label{fig:src_inside_0_4}
\end{figure}

\paragraph{Single-event analysis}
We inferred the possible regions of origin of the 40 events using the method discussed in Sec.~\ref{sec:backtracking}, as shown in Fig.~\ref{fig:CL}.
Sources compatible within the inferred regions were selected taking into account the horizon estimation discussed in Sec.~\ref{sec:horizon}. In Fig.~\ref{fig:src_inside_0_4}, summary plots for two events are shown. Here, the indication of the exclusion region due to the horizon is only used for illustrative purposes, as the exact determination has a level of arbitrariness.

For nearly all the 40 UHECR events, potential astrophysical sources, such as active galactic nuclei, starburst galaxies, and radio galaxies, are found within the identified volume. The only exception is the case of \texttt{PAO180812}, for which no such candidates were identified, except for those in the ``all galaxies'' and ``all galaxies B'' catalogs. However, these galaxies tend to have relatively low fluxes, making them weak candidate sources.
%
%However, galaxies from the 2MASS and the Biteau catalogs are present within the search region, although they tend to have relatively low fluxes, making them weaker candidate sources.
%

An important consideration is the $\sim$14\% systematic uncertainty in the energy scale. To account for this, we repeated the analysis also for the cases $E = {E_\text{nom} \pm 14\%}$. 
Under a $-14\%$ shift, the \texttt{PAO180812} event is compatible with two marginal ($\lesssim 5\%$ relative flux weights) AGNs traced by their X-ray emission.

Analyzing individual high-energy events does not provide enough statistical evidence to establish a definitive correlation with any specific class of astrophysical sources. A statistical analysis is needed to provide quantitative constraints.
%However, the case of the \texttt{PAO180812} event remains surely interesting, leaving alone systematic effects, potentially being interpreted as a contribution from \CE{ultra}-heavy elements beyond iron or as a transient \cite{Zhang:2024sjp, Farrar:2021kug}, or as the effect of a potentially non-negligible influence of the EGMF \cite{vanVliet:2021eah}, which was not included in our study.
However, the case of the \texttt{PAO180812} event remains interesting. Setting aside systematic effects and under the assumption of the GMF models we are considering, it could potentially be interpreted as the effect of a non-negligible influence of the EGMF \cite{Durrer:2013pga}, which was not included in our study, or as a transient, or perhaps as a contribution from ultra-heavy elements beyond iron \cite{Zhang:2024sjp}.

\paragraph{Likelihood analysis} The Galactic magnetic field maps the position of a source ($\hat{N}_{\text{src}}$) to the arrival direction of an event ($\hat{n}_{\text{evt}}$). Assuming a GMF model and a rigidity $\mathcal{R} = E/Z$, backtracking allows us to compute the inverse mapping $\hat{n}_{\text{evt}}$ $\rightarrow$ $\hat{N}_{\text{src}}$ in one shot, whereas the direct mapping is much less trivial to determine. The probability $p(\hat{N}_{\text{src}} \mid \hat{n}_{\text{evt}})$ can be estimated by backtracking $\hat{n}_{\text{evt}}$ over many realizations of the random field, detected energy, arrival direction, and particle charge, assuming a fixed model variation. This probability can in turn be used to approximate the inverse quantity, $p(\hat{n}_{\text{evt}} \mid \hat{N}_{\text{src}})$. 

\begin{figure}[H]
    \centering
    \includegraphics[width=0.7\linewidth]{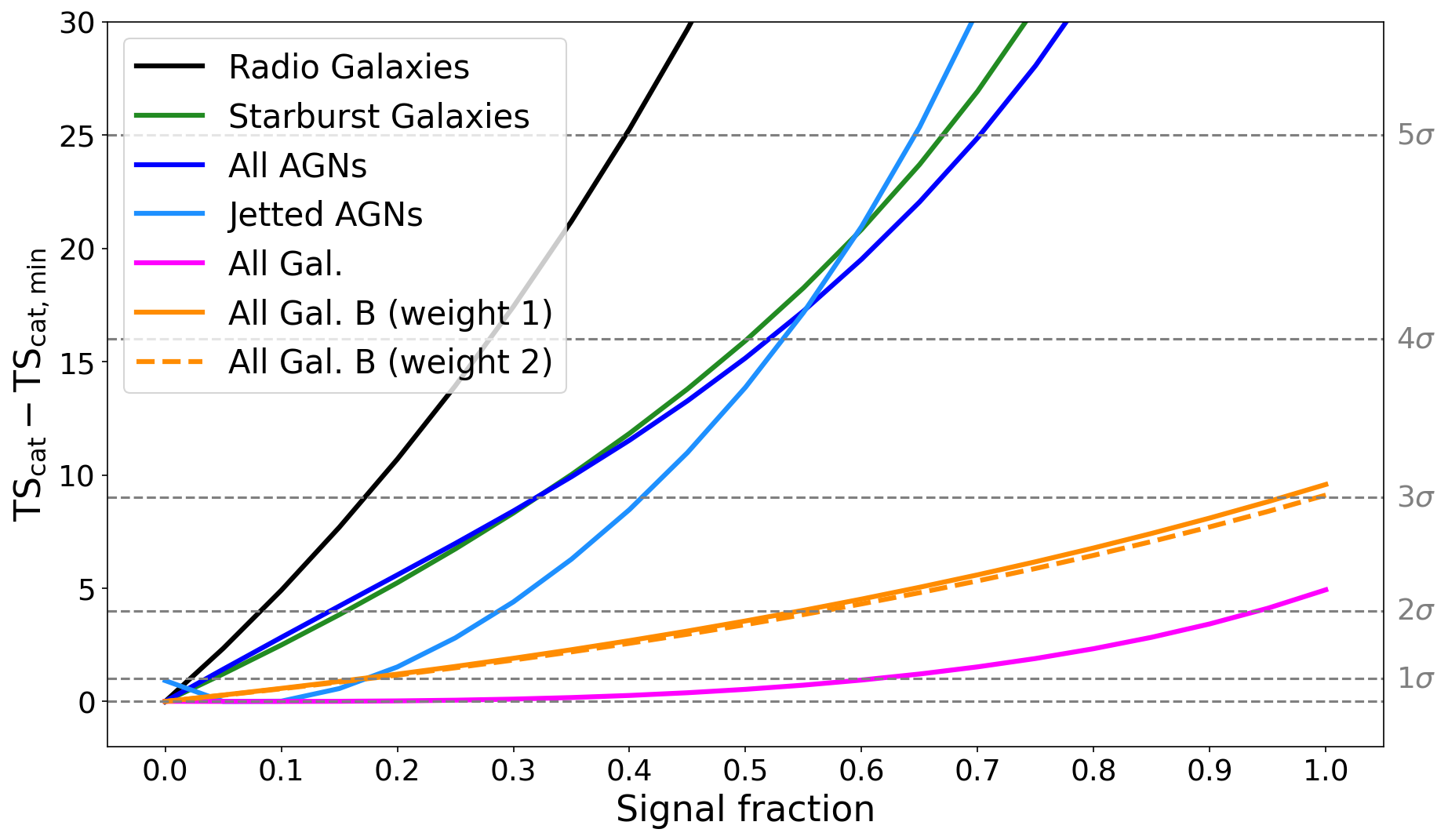}
    \includegraphics[width=0.7\linewidth]{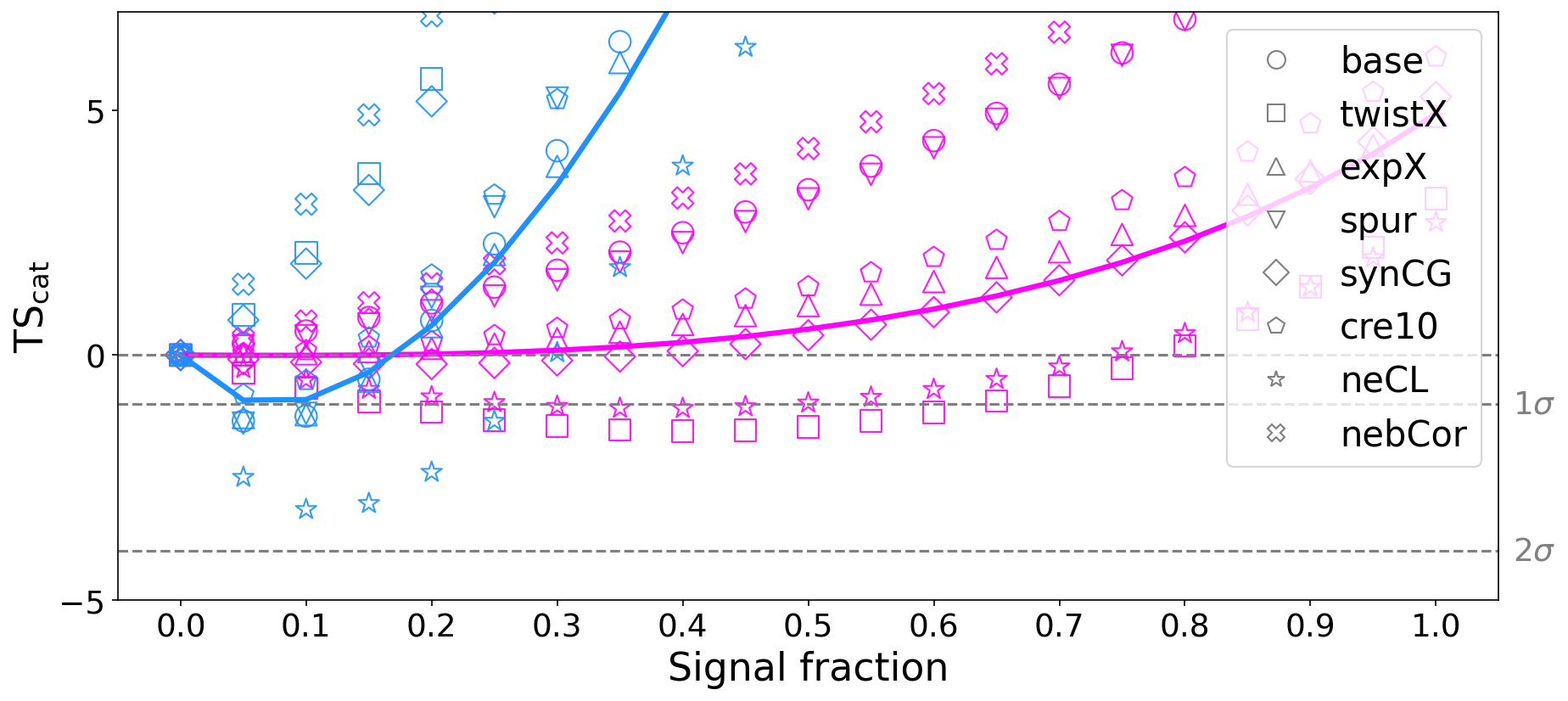}
    \caption{\small The test statistic (TS, see Eq.~\ref{eq:TS}) is shown as a function of the signal fraction ($1-f_{\text{iso}}$) for different source classes. The marginal TS over the different GMF models ($\exp(-\mathrm{TS}_\text{marg}/2) = \frac{1}{8}\sum_{i=1}^{8}\exp(-\mathrm{TS}_i/2)$) is computed for each catalog. \textbf{Top panel:} The quantity $\mathrm{TS}-\mathrm{TS_{min}}$, marginalized over GMF models, is shown as a solid line for each catalog. \textbf{Bottom panel:} The marginal TS is shown as a solid line, while the TS values for individual GMF models are represented with different marker styles. A limited number of catalogs are displayed to better appreciate the cases where the TS is negative and the model is preferred over isotropy.} 
    %The TS is shown as a function of the signal fraction.
    %This is computed separately for each GMF model in the UF23 ensemble and each source class over all the 40 highest-energy events detected by the Auger Observatory. The dashed line ($\mathrm{TS} = 25$) represents the $5\sigma$ level, while the continuous gray line represents the TS in the pure isotropic hypothesis.
    %Top panel: The marginal TS over the different GMF models ($\exp(-\mathrm{TS}_\text{marg}/2) = \frac{1}{8}\sum_{i=1}^{8}\exp(-\mathrm{TS}_i/2)$) is represented by a solid line for each catalog. Bottom panel: The TS is computed separately for each GMF model, each marker style representing a different model. The number of catalogs shown here is limited to better appreciate the cases where the TS is negative.}
    \label{fig:TS}
\end{figure}

Given a model, a source, and an event, the likelihood $\mathcal{L}_{\text{src},\text{evt},\text{model}}$ can be computed via a kernel density estimate (KDE) of the pixel density $\rho_j$ obtained as described in Sec.~\ref{sec:backtracking}. The KDE is smooth, therefore to each pixel (or source position) we can now assign a probability value.
For a source catalog, event, and model, the likelihood can then be expressed as:
\begin{equation}
\mathcal{L}_{\text{cat},\text{evt},\text{model}} = \sum_\text{src} w_{\text{src}} \mathcal{L}_{\text{src},\text{evt},\text{model}}\text{,}
\end{equation}
where  $w_{\text{src}}$ is a weight assigned to each source, computed as the product of its flux by the attenuation factor $a(E_i,d)$, defined in Eq.~(\ref{eq:attenuation_window}) and evaluated in the energy $E_i$ of the event and the distance $d$ of the source. This is normalized such that $\sum_{\text{src}} w_{\text{src}} = 1$. This weight is fundamental for our analysis, since it allows the three-dimensionality of our constraints. Additionally, for radio galaxies, the minimum power required for the acceleration is taken into account, as in \cite{Matthews:2018laz}.
% and evaluated in the energy $E_i$ of the event and the distance $d$ of each source
%
Moreover, an isotropic fraction can be added to the equation:
\begin{equation}
    \mathcal{L}_{\text{cat},\text{evt},\text{model}} = (1-f_\text{iso}) \mathcal{L}_{\text{cat},\text{evt},\text{model}} + f_\text{iso},
    \end{equation}
the isotropic likelihood being $\mathcal{L}_\text{iso}=1$ due to our choice of normalization.
Finally, assuming statistical independence among the observed events, the total likelihood for the model is given by the product over all events,
%,     $\mathcal{L}_{\text{cat, model}} = \Pi_\text{evt} \mathcal{L}_{\text{cat,evt},\text{model}}$,
and can be evaluated in logarithmic form for convenience, as $\ln \mathcal{L}_\text{cat,model} = \sum_\text{evt} \ln \mathcal{L}_{\text{cat,evt},\text{model}}$.
% \begin{equation}
%     \ln \mathcal{L}_\text{cat,model} = \sum_\text{evt} \ln \mathcal{L}_{\text{cat,evt},\text{model}}.
%     \label{eq:lnL}
% \end{equation}
The total likelihood can then be converted into a test statistic (TS), as:
\begin{equation}
     \mathrm{TS_{\text{cat,model}}} = -2\ln\mathcal{L}_{\text{cat,model}}.
    \label{eq:TS}
\end{equation}
% where $\sqrt{\mathrm{TS}}$ represents a significance, while $\sqrt{\mathrm{TS}-\mathrm{TS_{min}}}$ represents a pseudo-significance.
%
% \TB{The signal fraction $(1-f_\text{iso})$} is scanned from 0 to 1 in steps of 0.05. By construction, the pure isotropic case corresponds to $\mathrm{TS}=0$.
% The results of this analysis are presented in Fig.~\ref{fig:TS}.
The signal fraction $(1-f_\text{iso})$ is scanned from $0$ to $1$ in steps of $0.05$. By construction, the pure isotropic case corresponds to $\mathrm{TS}=0$. The significance at which a given signal fraction is disfavored with respect to the best-fit
one (minimum TS, $\mathrm{TS}_{\min}$) is given by~$\sqrt{\mathrm{TS} -
\mathrm{TS}_{\min}}$. In particular, for the catalogs where the minimum TS is~0 at~$f_\text{iso} = 1$,  $\sqrt{\mathrm{TS}}$~is
the significance at which a given signal fraction is disfavored
with respect to isotropy.
The results of this analysis are presented in Fig.~\ref{fig:TS}.

Starburst and radio galaxies, and active galactic nuclei can be excluded as dominant contributors to the observed cosmic-ray flux above 100~EeV at the $5\sigma$ level.
In particular, we exclude signal fractions greater than 40\%, 65\%, 67\% and 70\% for radio galaxies, jetted AGNs traced by their $\gamma$-ray emission, starburst galaxies, and AGNs traced by their X-ray emission, respectively.
The ``jetted AGNs'' catalog provides a better description of the data than the isotropic case, when small signal fractions are considered, while the ``all galaxies'' catalog does only in a few specific GMF configurations, at intermediate signal fractions.
The most significant scenario consists in a 10\% signal fraction of jetted AGN sources in the \texttt{neCL} GMF model.

% In conclusion, no single catalog of starburst, radio galaxies, and active galactic nuclei can account for a large fraction of the observed events above 100~EeV.
% This suggests that more than one type of source populations should contribute, or that more common galaxies, such as those included in the 2MASS catalog, may also play a role. Alternatively, the effects of extra-galactic magnetic fields might sizeably deflect the UHECRs before they arrive to our Galaxy, or a contribution from ultra-heavy nuclei could be relevant as well.

\section{Conclusions}
\label{sec:conclusions}

In this work, we reconstructed the possible sky regions of origin of the highest-energy cosmic rays detected by the Pierre Auger Observatory during Phase~I through Galactic backtracking, including constraints on the maximum propagation distance.
Two complementary approaches were presented: a single-event analysis and a likelihood-based framework. 

%The single-event analysis identified at least one plausible astrophysical source within the localization region for the majority of events, with the exception of \texttt{PAO180812}, which exhibits no clear counterpart among known source classes.
%
 %The nature of this outlier remains uncertain; it could be linked to a transient event, indicate a non-negligible influence of the extragalactic magnetic field, or suggest the presence of super-heavy nuclei.
%
%This outlier may indicate the presence of super-heavy nuclei, hence a contribution from transients, or may suggest a non-negligible influence of the extragalactic magnetic field. 
%However, it is worth to mention that also the systematic of the Auger energy scale should be taken into account, since this could contribute to a different result.

From the single-event analysis, we find that every event but one has at least one plausible astrophysical counterpart within the $95\%$ CL localization region. The sole exception is \texttt{PAO180812}, which exhibits no clear counterpart among known  source classes -- but does so when the 14\% systematic energy shift is accounted for.
%The sole exception, \texttt{PAO180812}, shows no convincing association: even after accounting for the $\pm14\%$ systematic uncertainty on the Auger energy scale we identify only two marginal ($\lesssim 5\%$ relative flux weights) candidates in a single catalog.
%Its origin therefore remains enigmatic, possibly pointing to non-negligible EGMFs, a transient accelerator, or ultra-heavy nuclear primaries.

% \GG{The single-event analysis identified at least one plausible astrophysical source within the localization region for the majority of events, with the exception of PAO180812, which exhibits no clear counterpart among known source classes (but does so when the 14\% systematic energy shift is accounted for).}

The likelihood analysis indicates that, within the set of assumptions and the approach adopted here, most tested source catalogs can be excluded as dominant contributors to the ultra-high-energy cosmic ray flux above 100~EeV. % at the $5\sigma$ confidence level.
This may suggest that more than one type of source populations should contribute, or that more common galaxies could also play a role. Also, the EGMF might significantly deflect the UHECRs before they arrive to our Galaxy, or a contribution from ultra-heavy nuclei could also be relevant. 

%, with the exception of the 2MASS and Biteau catalogs.
% The Fermi and the 2MASS catalogs exhibit a modest preference over isotropy at low signal fractions. The most favorable configuration is found from a mixed-source model combining the 2MASS ($f = 20\%$) and the Fermi ($f = 10\%$) contributions.

% \begingroup
% \setlength{\bibsep}{0pt}
% \footnotesize
% \bibliographystyle{unsrt}
% \bibliography{bib}
% \endgroup

\begingroup
\setlength{\bibsep}{0.5pt}
\small
\bibliographystyle{unsrt}
\bibliography{bib}
\endgroup

\newpage

\clearpage
\section*{The Pierre Auger Collaboration}

{\footnotesize\setlength{\baselineskip}{10pt}
\noindent
\begin{wrapfigure}[11]{l}{0.12\linewidth}
\vspace{-4pt}
\includegraphics[width=0.98\linewidth]{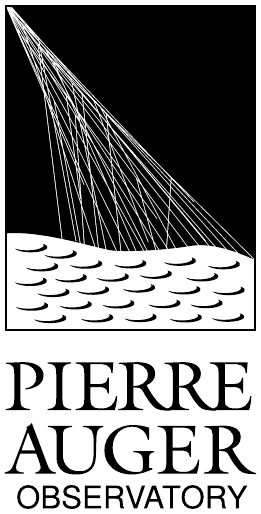}
\end{wrapfigure}
\begin{sloppypar}\noindent
% created on 2025-06-06
A.~Abdul Halim$^{13}$,
P.~Abreu$^{70}$,
M.~Aglietta$^{53,51}$,
I.~Allekotte$^{1}$,
K.~Almeida Cheminant$^{78,77}$,
A.~Almela$^{7,12}$,
R.~Aloisio$^{44,45}$,
J.~Alvarez-Mu\~niz$^{76}$,
A.~Ambrosone$^{44}$,
J.~Ammerman Yebra$^{76}$
% ...
\end{sloppypar}
\begin{center}
\end{center}

\vspace{1ex}
% created on 2025-06-06
% needs \usepackage{enumitem}
\begin{description}[labelsep=0.2em,align=right,labelwidth=0.7em,labelindent=0em,leftmargin=2em,noitemsep,before={\renewcommand\makelabel[1]{##1 }}]
\item[$^{1}$] Centro At\'o{}mico Bariloche and Instituto Balseiro (CNEA-UNCuyo-CONICET), San Carlos de Bariloche, Argentina
\item[$^{2}$] Departamento de F\'\i{}sica and Departamento de Ciencias de la Atm\'osfera y los Oc\'eanos, FCEyN, Universidad de Buenos Aires and CONICET, Buenos Aires, Argentina
\item[$^{3}$] IFLP, Universidad Nacional de La Plata and CONICET, La Plata, Argentina
\item[$^{4}$] Instituto de Astronom\'\i{}a y F\'\i{}sica del Espacio (IAFE, CONICET-UBA), Buenos Aires, Argentina
\item[$^{5}$] Instituto de F\'\i{}sica de Rosario (IFIR) -- CONICET/U.N.R.\ and Facultad de Ciencias Bioqu\'\i{}micas y Farmac\'euticas U.N.R., Rosario, Argentina
\item[$^{6}$] Instituto de Tecnolog\'\i{}as en Detecci\'on y Astropart\'\i{}culas (CNEA, CONICET, UNSAM), and Universidad Tecnol\'ogica Nacional -- Facultad Regional Mendoza (CONICET/CNEA), Mendoza, Argentina
\item[$^{7}$] Instituto de Tecnolog\'\i{}as en Detecci\'on y Astropart\'\i{}culas (CNEA, CONICET, UNSAM), Buenos Aires, Argentina
\item[$^{8}$] International Center of Advanced Studies and Instituto de Ciencias F\'\i{}sicas, ECyT-UNSAM and CONICET, Campus Miguelete -- San Mart\'\i{}n, Buenos Aires, Argentina
\item[$^{9}$] Laboratorio Atm\'osfera -- Departamento de Investigaciones en L\'aseres y sus Aplicaciones -- UNIDEF (CITEDEF-CONICET), Argentina
\item[$^{10}$] Observatorio Pierre Auger, Malarg\"ue, Argentina
\item[$^{11}$] Observatorio Pierre Auger and Comisi\'on Nacional de Energ\'\i{}a At\'omica, Malarg\"ue, Argentina
\item[$^{12}$] Universidad Tecnol\'ogica Nacional -- Facultad Regional Buenos Aires, Buenos Aires, Argentina
\item[$^{13}$] University of Adelaide, Adelaide, S.A., Australia
\item[$^{14}$] Universit\'e Libre de Bruxelles (ULB), Brussels, Belgium
\item[$^{15}$] Vrije Universiteit Brussels, Brussels, Belgium
\item[$^{16}$] Centro Brasileiro de Pesquisas Fisicas, Rio de Janeiro, RJ, Brazil
\item[$^{17}$] Centro Federal de Educa\c{c}\~ao Tecnol\'ogica Celso Suckow da Fonseca, Petropolis, Brazil
\item[$^{18}$] Instituto Federal de Educa\c{c}\~ao, Ci\^encia e Tecnologia do Rio de Janeiro (IFRJ), Brazil
\item[$^{19}$] Universidade de S\~ao Paulo, Escola de Engenharia de Lorena, Lorena, SP, Brazil
\item[$^{20}$] Universidade de S\~ao Paulo, Instituto de F\'\i{}sica de S\~ao Carlos, S\~ao Carlos, SP, Brazil
\item[$^{21}$] Universidade de S\~ao Paulo, Instituto de F\'\i{}sica, S\~ao Paulo, SP, Brazil
\item[$^{22}$] Universidade Estadual de Campinas (UNICAMP), IFGW, Campinas, SP, Brazil
\item[$^{23}$] Universidade Estadual de Feira de Santana, Feira de Santana, Brazil
\item[$^{24}$] Universidade Federal de Campina Grande, Centro de Ciencias e Tecnologia, Campina Grande, Brazil
\item[$^{25}$] Universidade Federal do ABC, Santo Andr\'e, SP, Brazil
\item[$^{26}$] Universidade Federal do Paran\'a, Setor Palotina, Palotina, Brazil
\item[$^{27}$] Universidade Federal do Rio de Janeiro, Instituto de F\'\i{}sica, Rio de Janeiro, RJ, Brazil
\item[$^{28}$] Universidad de Medell\'\i{}n, Medell\'\i{}n, Colombia
\item[$^{29}$] Universidad Industrial de Santander, Bucaramanga, Colombia
\item[$^{30}$] Charles University, Faculty of Mathematics and Physics, Institute of Particle and Nuclear Physics, Prague, Czech Republic
\item[$^{31}$] Institute of Physics of the Czech Academy of Sciences, Prague, Czech Republic
\item[$^{32}$] Palacky University, Olomouc, Czech Republic
\item[$^{33}$] CNRS/IN2P3, IJCLab, Universit\'e Paris-Saclay, Orsay, France
\item[$^{34}$] Laboratoire de Physique Nucl\'eaire et de Hautes Energies (LPNHE), Sorbonne Universit\'e, Universit\'e de Paris, CNRS-IN2P3, Paris, France
\item[$^{35}$] Univ.\ Grenoble Alpes, CNRS, Grenoble Institute of Engineering Univ.\ Grenoble Alpes, LPSC-IN2P3, 38000 Grenoble, France
\item[$^{36}$] Universit\'e Paris-Saclay, CNRS/IN2P3, IJCLab, Orsay, France
\item[$^{37}$] Bergische Universit\"at Wuppertal, Department of Physics, Wuppertal, Germany
\item[$^{38}$] Karlsruhe Institute of Technology (KIT), Institute for Experimental Particle Physics, Karlsruhe, Germany
\item[$^{39}$] Karlsruhe Institute of Technology (KIT), Institut f\"ur Prozessdatenverarbeitung und Elektronik, Karlsruhe, Germany
\item[$^{40}$] Karlsruhe Institute of Technology (KIT), Institute for Astroparticle Physics, Karlsruhe, Germany
\item[$^{41}$] RWTH Aachen University, III.\ Physikalisches Institut A, Aachen, Germany
\item[$^{42}$] Universit\"at Hamburg, II.\ Institut f\"ur Theoretische Physik, Hamburg, Germany
\item[$^{43}$] Universit\"at Siegen, Department Physik -- Experimentelle Teilchenphysik, Siegen, Germany
\item[$^{44}$] Gran Sasso Science Institute, L'Aquila, Italy
\item[$^{45}$] INFN Laboratori Nazionali del Gran Sasso, Assergi (L'Aquila), Italy
\item[$^{46}$] INFN, Sezione di Catania, Catania, Italy
\item[$^{47}$] INFN, Sezione di Lecce, Lecce, Italy
\item[$^{48}$] INFN, Sezione di Milano, Milano, Italy
\item[$^{49}$] INFN, Sezione di Napoli, Napoli, Italy
\item[$^{50}$] INFN, Sezione di Roma ``Tor Vergata'', Roma, Italy
\item[$^{51}$] INFN, Sezione di Torino, Torino, Italy
\item[$^{52}$] Istituto di Astrofisica Spaziale e Fisica Cosmica di Palermo (INAF), Palermo, Italy
\item[$^{53}$] Osservatorio Astrofisico di Torino (INAF), Torino, Italy
\item[$^{54}$] Politecnico di Milano, Dipartimento di Scienze e Tecnologie Aerospaziali , Milano, Italy
\item[$^{55}$] Universit\`a del Salento, Dipartimento di Matematica e Fisica ``E.\ De Giorgi'', Lecce, Italy
\item[$^{56}$] Universit\`a dell'Aquila, Dipartimento di Scienze Fisiche e Chimiche, L'Aquila, Italy
\item[$^{57}$] Universit\`a di Catania, Dipartimento di Fisica e Astronomia ``Ettore Majorana``, Catania, Italy
\item[$^{58}$] Universit\`a di Milano, Dipartimento di Fisica, Milano, Italy
\item[$^{59}$] Universit\`a di Napoli ``Federico II'', Dipartimento di Fisica ``Ettore Pancini'', Napoli, Italy
\item[$^{60}$] Universit\`a di Palermo, Dipartimento di Fisica e Chimica ''E.\ Segr\`e'', Palermo, Italy
\item[$^{61}$] Universit\`a di Roma ``Tor Vergata'', Dipartimento di Fisica, Roma, Italy
\item[$^{62}$] Universit\`a Torino, Dipartimento di Fisica, Torino, Italy
\item[$^{63}$] Benem\'erita Universidad Aut\'onoma de Puebla, Puebla, M\'exico
\item[$^{64}$] Unidad Profesional Interdisciplinaria en Ingenier\'\i{}a y Tecnolog\'\i{}as Avanzadas del Instituto Polit\'ecnico Nacional (UPIITA-IPN), M\'exico, D.F., M\'exico
\item[$^{65}$] Universidad Aut\'onoma de Chiapas, Tuxtla Guti\'errez, Chiapas, M\'exico
\item[$^{66}$] Universidad Michoacana de San Nicol\'as de Hidalgo, Morelia, Michoac\'an, M\'exico
\item[$^{67}$] Universidad Nacional Aut\'onoma de M\'exico, M\'exico, D.F., M\'exico
\item[$^{68}$] Institute of Nuclear Physics PAN, Krakow, Poland
\item[$^{69}$] University of \L{}\'od\'z, Faculty of High-Energy Astrophysics,\L{}\'od\'z, Poland
\item[$^{70}$] Laborat\'orio de Instrumenta\c{c}\~ao e F\'\i{}sica Experimental de Part\'\i{}culas -- LIP and Instituto Superior T\'ecnico -- IST, Universidade de Lisboa -- UL, Lisboa, Portugal
\item[$^{71}$] ``Horia Hulubei'' National Institute for Physics and Nuclear Engineering, Bucharest-Magurele, Romania
\item[$^{72}$] Institute of Space Science, Bucharest-Magurele, Romania
\item[$^{73}$] Center for Astrophysics and Cosmology (CAC), University of Nova Gorica, Nova Gorica, Slovenia
\item[$^{74}$] Experimental Particle Physics Department, J.\ Stefan Institute, Ljubljana, Slovenia
\item[$^{75}$] Universidad de Granada and C.A.F.P.E., Granada, Spain
\item[$^{76}$] Instituto Galego de F\'\i{}sica de Altas Enerx\'\i{}as (IGFAE), Universidade de Santiago de Compostela, Santiago de Compostela, Spain
\item[$^{77}$] IMAPP, Radboud University Nijmegen, Nijmegen, The Netherlands
\item[$^{78}$] Nationaal Instituut voor Kernfysica en Hoge Energie Fysica (NIKHEF), Science Park, Amsterdam, The Netherlands
\item[$^{79}$] Stichting Astronomisch Onderzoek in Nederland (ASTRON), Dwingeloo, The Netherlands
\item[$^{80}$] Universiteit van Amsterdam, Faculty of Science, Amsterdam, The Netherlands
\item[$^{81}$] Case Western Reserve University, Cleveland, OH, USA
\item[$^{82}$] Colorado School of Mines, Golden, CO, USA
\item[$^{83}$] Department of Physics and Astronomy, Lehman College, City University of New York, Bronx, NY, USA
\item[$^{84}$] Michigan Technological University, Houghton, MI, USA
\item[$^{85}$] New York University, New York, NY, USA
\item[$^{86}$] University of Chicago, Enrico Fermi Institute, Chicago, IL, USA
\item[$^{87}$] University of Delaware, Department of Physics and Astronomy, Bartol Research Institute, Newark, DE, USA
\item[] -----
\item[$^{a}$] Max-Planck-Institut f\"ur Radioastronomie, Bonn, Germany
\item[$^{b}$] also at Kapteyn Institute, University of Groningen, Groningen, The Netherlands
\item[$^{c}$] School of Physics and Astronomy, University of Leeds, Leeds, United Kingdom
\item[$^{d}$] Fermi National Accelerator Laboratory, Fermilab, Batavia, IL, USA
\item[$^{e}$] Pennsylvania State University, University Park, PA, USA
\item[$^{f}$] Colorado State University, Fort Collins, CO, USA
\item[$^{g}$] Louisiana State University, Baton Rouge, LA, USA
\item[$^{h}$] now at Graduate School of Science, Osaka Metropolitan University, Osaka, Japan
\item[$^{i}$] Institut universitaire de France (IUF), France
\item[$^{j}$] now at Technische Universit\"at Dortmund and Ruhr-Universit\"at Bochum, Dortmund and Bochum, Germany
\end{description}

% created on 2025-06-06
\section*{Acknowledgments}

\begin{sloppypar}
The successful installation, commissioning, and operation of the Pierre
Auger Observatory would not have been possible without the strong
commitment and effort from the technical and administrative staff in
Malarg\"ue. We are very grateful to the following agencies and
organizations for financial support:
\end{sloppypar}

\begin{sloppypar}
Argentina -- Comisi\'on Nacional de Energ\'\i{}a At\'omica; Agencia Nacional de
Promoci\'on Cient\'\i{}fica y Tecnol\'ogica (ANPCyT); Consejo Nacional de
Investigaciones Cient\'\i{}ficas y T\'ecnicas (CONICET); Gobierno de la
Provincia de Mendoza; Municipalidad de Malarg\"ue; NDM Holdings and Valle
Las Le\~nas; in gratitude for their continuing cooperation over land
access; Australia -- the Australian Research Council; Belgium -- Fonds
de la Recherche Scientifique (FNRS); Research Foundation Flanders (FWO),
Marie Curie Action of the European Union Grant No.~101107047; Brazil --
Conselho Nacional de Desenvolvimento Cient\'\i{}fico e Tecnol\'ogico (CNPq);
Financiadora de Estudos e Projetos (FINEP); Funda\c{c}\~ao de Amparo \`a
Pesquisa do Estado de Rio de Janeiro (FAPERJ); S\~ao Paulo Research
Foundation (FAPESP) Grants No.~2019/10151-2, No.~2010/07359-6 and
No.~1999/05404-3; Minist\'erio da Ci\^encia, Tecnologia, Inova\c{c}\~oes e
Comunica\c{c}\~oes (MCTIC); Czech Republic -- GACR 24-13049S, CAS LQ100102401,
MEYS LM2023032, CZ.02.1.01/0.0/0.0/16{\textunderscore}013/0001402,
CZ.02.1.01/0.0/0.0/18{\textunderscore}046/0016010 and
CZ.02.1.01/0.0/0.0/17{\textunderscore}049/0008422 and CZ.02.01.01/00/22{\textunderscore}008/0004632;
France -- Centre de Calcul IN2P3/CNRS; Centre National de la Recherche
Scientifique (CNRS); Conseil R\'egional Ile-de-France; D\'epartement
Physique Nucl\'eaire et Corpusculaire (PNC-IN2P3/CNRS); D\'epartement
Sciences de l'Univers (SDU-INSU/CNRS); Institut Lagrange de Paris (ILP)
Grant No.~LABEX ANR-10-LABX-63 within the Investissements d'Avenir
Programme Grant No.~ANR-11-IDEX-0004-02; Germany -- Bundesministerium
f\"ur Bildung und Forschung (BMBF); Deutsche Forschungsgemeinschaft (DFG);
Finanzministerium Baden-W\"urttemberg; Helmholtz Alliance for
Astroparticle Physics (HAP); Helmholtz-Gemeinschaft Deutscher
Forschungszentren (HGF); Ministerium f\"ur Kultur und Wissenschaft des
Landes Nordrhein-Westfalen; Ministerium f\"ur Wissenschaft, Forschung und
Kunst des Landes Baden-W\"urttemberg; Italy -- Istituto Nazionale di
Fisica Nucleare (INFN); Istituto Nazionale di Astrofisica (INAF);
Ministero dell'Universit\`a e della Ricerca (MUR); CETEMPS Center of
Excellence; Ministero degli Affari Esteri (MAE), ICSC Centro Nazionale
di Ricerca in High Performance Computing, Big Data and Quantum
Computing, funded by European Union NextGenerationEU, reference code
CN{\textunderscore}00000013; M\'exico -- Consejo Nacional de Ciencia y Tecnolog\'\i{}a
(CONACYT) No.~167733; Universidad Nacional Aut\'onoma de M\'exico (UNAM);
PAPIIT DGAPA-UNAM; The Netherlands -- Ministry of Education, Culture and
Science; Netherlands Organisation for Scientific Research (NWO); Dutch
national e-infrastructure with the support of SURF Cooperative; Poland
-- Ministry of Education and Science, grants No.~DIR/WK/2018/11 and
2022/WK/12; National Science Centre, grants No.~2016/22/M/ST9/00198,
2016/23/B/ST9/01635, 2020/39/B/ST9/01398, and 2022/45/B/ST9/02163;
Portugal -- Portuguese national funds and FEDER funds within Programa
Operacional Factores de Competitividade through Funda\c{c}\~ao para a Ci\^encia
e a Tecnologia (COMPETE); Romania -- Ministry of Research, Innovation
and Digitization, CNCS-UEFISCDI, contract no.~30N/2023 under Romanian
National Core Program LAPLAS VII, grant no.~PN 23 21 01 02 and project
number PN-III-P1-1.1-TE-2021-0924/TE57/2022, within PNCDI III; Slovenia
-- Slovenian Research Agency, grants P1-0031, P1-0385, I0-0033, N1-0111;
Spain -- Ministerio de Ciencia e Innovaci\'on/Agencia Estatal de
Investigaci\'on (PID2019-105544GB-I00, PID2022-140510NB-I00 and
RYC2019-027017-I), Xunta de Galicia (CIGUS Network of Research Centers,
Consolidaci\'on 2021 GRC GI-2033, ED431C-2021/22 and ED431F-2022/15),
Junta de Andaluc\'\i{}a (SOMM17/6104/UGR and P18-FR-4314), and the European
Union (Marie Sklodowska-Curie 101065027 and ERDF); USA -- Department of
Energy, Contracts No.~DE-AC02-07CH11359, No.~DE-FR02-04ER41300,
No.~DE-FG02-99ER41107 and No.~DE-SC0011689; National Science Foundation,
Grant No.~0450696, and NSF-2013199; The Grainger Foundation; Marie
Curie-IRSES/EPLANET; European Particle Physics Latin American Network;
and UNESCO.
\end{sloppypar}

}

\end{document}